\documentstyle[preprint,tighten,prl,aps,epsfig,epsf]{revtex}

\begin{document}


\title{Growth and spin-resolved photoelectron spectroscopy\\ of the MgO/Fe(110) system}

\author{Yu. S. Dedkov, M. Fonin, and G. G\"untherodt}
\address{II. Physikalisches Institut, Rheinisch-Westf\"alische Technische
         Hochschule Aachen,\\ 52056 Aachen, Germany}
\author{U. R\"udiger}
\address{Fachbereich Physik, Universit\"at Konstanz, D-78457,
Konstanz, Germany}

\maketitle


\begin{abstract}
Electronic and structural properties of ultrathin MgO layers grown
on epitaxial Fe(110) films were investigated at room temperature
by means of spin-resolved photoelectron spectroscopy,
Auger-electron spectroscopy, and low energy electron diffraction.
The spin polarization at the Fermi level of the Fe(110) film
decreases abruptly with increasing thickness of MgO layer up to
7\,\AA. This behavior is attributed to the formation of a thin FeO
layer at the MgO(111)/Fe(110) interface, attenuating the intrinsic
spin polarization.
\end{abstract}

\pagebreak The high tunneling magnetoresistance (TMR) values
achievable by means of magnetic tunnel junctions (MTJs) consisting
of two ferromagnetic electrodes separated by a thin insulating
layer have attracted strong interest for potential applications in
magnetoelectronics~\cite{Gallagher:1997,Moodera:1995,Parkin:1999,Boeve:2001}.
According to Julli$\grave{e}$re's model magnetoresistance of such
MTJs depends only on the spin polarization of the ferromagnetic
electrodes used. In contrast, \textit{ab initio} electronic
structure and transport calculations have shown that the
magnetoelectronic properties of such devices strongly depend on
the structural as well as electronic properties of the insulating
layer and the specific termination at the insulator/ferromagnet
(I/FM)
interface~\cite{MacLaren:1999,Mavropoulos:2000,Oleinik:2000,Butler:2001}.

In the last few years the epitaxial Fe/MgO/Fe(100) MTJ system has
been intensively studied. For such ideal MTJs with abrupt
interfaces between MgO and Fe, the TMR values are predicted to be
as high as $\sim$2000\%~\cite{Butler:2001}. Recent experiments on
the Fe(100)/MgO/FeCo MTJ~\cite{Bowen:2001} and Fe/MgO/Fe
MTJ~\cite{Faure:2003} show a TMR value of only 60\% at 30\,K and
100\% at 80\,K, respectively, which are close to the theoretically
predicted value of 75\% for the system assuming a FeO layer at the
MgO/Fe interface~\cite{Butler:unp}. This fact can be considered as
an indirect proof for the iron oxide formation at the
inhomogeneous MgO/Fe interface. The surface x-ray diffraction
(SXRD) and vibration spectroscopy
experiments~\cite{Meyerheim:2001,Meyerheim:2002,Oh:2003} which
were carried out on the MgO/Fe(100) system gave a direct evidence
for the formation of a FeO sub-monolayer at the interface.

In the present work the crystallographic and electronic structure
of the MgO/Fe(110) interface were investigated by means of low
energy electron diffraction (LEED), Auger electron spectroscopy
(AES), and spin- and angle-resolved photoelectron spectroscopy
(SPARPES). For this purpose MgO layers with (111) orientation can
be epitaxially grown on the \textit{bcc} \textit{d}-metal film as
it was shown earlier~\cite{Purnell:1994,Magkoev:2002}. A TMR
effect of about 30\% with Fe(110) layers as magnetic electrodes
and amorphous Al$_2$O$_3$ layer as insulator was demonstrated by
Yuasa \textit{et al.}~\cite{Yuasa:2000}. The present SPARPES
experiments show that the spin polarization of photoelectrons at
the Fermi level (E$_{F}$) decreased rapidly with increasing MgO
layer thickness which can not only be due to the spin scattering
of photoelectrons in the MgO layer. This fact can be attributed to
the formation of a thin depolarizing FeO layer at the MgO/Fe
interface on the basis of a spin-dependent transport model of iron
valence band photoelectrons through the oxide overlayer.

All experiments were carried out at room temperature in a UHV
system for energy-resolved SPARPES with spin analysis described in
detail in Ref.~\cite{Raue:1984}. It consists of a UHV chamber
equipped with LEED optics, gas inlet, and an AES spectrometer with
a cylindrical mirror analyzer. All AES spectra were recorded in
the $dN/dE$ mode with 2.5\,keV primary electron energy and
peak-to-peak modulation voltage of 2\,V. The SPARPES spectra
(He\,I, h$\nu$=21.2\,eV) were recorded in normal emission by a
180$^\circ$ hemi-spherical energy analyzer connected to a 100 kV
Mott detector for spin analysis. The energy resolution is 100\,meV
and the angle resolution $\pm1^\circ$. The spin-resolved
measurements have been performed in magnetic remanence after
having applied a magnetic field pulse of about 500\,Oe along the
in-plane $<1{\bar1}0>$ easy axis of the thin Fe(110)
films~\cite{Kurzawa:1986}.

Clean 50-\AA\ thick Fe(110) films were prepared \textit{in situ}
by electron-beam evaporation onto a W(110) substrate, while the
thickness was simultaneously monitored by a quartz microbalance.
The degree of crystalline order of the thin epitaxial Fe(110)
films and the MgO overlayer on top of Fe(110) films was checked by
LEED. The surface cleanliness has been monitored by AES and
valence band PES. The MgO was deposited \textit{in situ} by
electron beam evaporation from a W-crucible. The base pressure in
the vacuum chamber was $1\times10^{-10}$\,mbar and increased up to
$1\times10^{-9}$\,mbar during the MgO deposition process.

Fig.1 shows LEED images of (a) the clean Fe(110) film as well as
(b) the 30-\AA\ thick MgO layer on top of the Fe(110) film. The
well-ordered hexagonal (1$\times$1) LEED structure of the MgO
layer is clearly visible. For intermediate thicknesses of MgO
(less than 30\,\AA) on top of the Fe(110) film the LEED images do
not show any well-ordered structure, which means that a smooth
transition from the \textit{bcc} (1$\times$1) structure to the
\textit{fcc} (1$\times$1) structure takes place. The spots in
these cases are very weak or not visible, with a diffuse
background (not shown here).

The modification of the electronic structure of the MgO/Fe(110)
system as function of the thickness of the deposited MgO layer was
investigated by means of photoelectron spectroscopy (PES) in
normal emission. The PES spectra obtained in this experiment are
shown in Fig.2. They show a pronounced structure corresponding to
the emission from the Fe (3\textit{d}) states in the range of
binding energies near E$_F$ as well as from the O (2\textit{p})
states in the range of 4-10\,eV. With an increasing amount of
deposited MgO a gradual shift of the maxima of the O (2\textit{p})
states and the valence band edge (VBE) toward of larger binding
energies can be clearly seen in Fig.2. In the inset of Fig.2 the
change of the onset of the VBE position is shown with an
increasing MgO layer thickness. This shift is approximately 1\,eV.
As it was discussed earlier by Kiguchi \textit{et
al.}~\cite{Kiguchi:2002} for the MgO/Ag(001) system this effect
can be attributed to the increase of the binding energy of the Mg
(3\textit{s}) states above E$_F$ and the increase of the binding
energy of the O (2\textit{p}) states, which results from the
change in the Madelung potential for thin films.

The spin-resolved electronic structure of the MgO/Fe(110) system
was studied by SPARPES. The spin-resolved photoemission spectra
together with the total emission intensity and the spin
polarization as a function of the binding energy of Fe(110),
2\,\AA\,MgO/Fe(110) as well as 5\,\AA\,MgO/Fe(110) are presented
in Fig.3a) (from bottom to top) and b), respectively. The
spin-resolved spectra of the valence band of the Fe(110) film show
the emission from the $\sum^1{\downarrow}\oplus
\sum^3{\downarrow}$ states near 0.25\,eV and from the
$\sum^1{\uparrow}\oplus \sum^4{\uparrow}$ states near 0.7\,eV. The
value of the spin polarization of about ($-80\pm5\%$) and the
shape of the spectra are in agreement with previous
measurements~\cite{Kurzawa:1986}. After the deposition of a 2-\AA\
thick MgO layer on the Fe(110) film surface the total intensity of
the photoemission spectra measured near E$_F$ decreases
drastically. At the same time, the features of the valence band of
Fe can still be observed. For this system the spin polarization
near E$_F$ is decreased to about $-(52\pm5)\%$ compared to
$-(80\pm5)\%$ of the clean Fe(110) surface. Additional deposition
of MgO on top of the Fe(110) film leads to a further decrease of
the spin polarization at E$_F$ (shown for the 5\,\AA\,MgO/Fe(110)
system in Fig.3 with $P(E_F)=-(21\pm5)\%$).

Fig.4 shows experimentally determined changes of the normalized
spin polarization at E$_F$ of the MgO/Fe(110) system as function
of the deposited MgO layer thickness. In general, the changes of
the spin polarization at $E_F$ of the emitted photoelectrons in
the system FM/oxide can be presented by the formula:

$P=\frac{J^+_0\exp(-\sigma^+{\cdot}\,d)-J^-_0\exp(-\sigma^-{\cdot}\,d)}{J^+_0\exp(-\sigma^+{\cdot}\,d)+J^-_0\exp(-\sigma^-{\cdot}\,d)}$,

where $J^+_0$($J^-_0$) is the spin-up (down) photoelectron
intensity without the oxide, $d$ is the thickness of the oxide,
and $\sigma^+$($\sigma^-$) is the scattering cross section for
spin-up (down) electrons ($\sigma=\frac{\sigma^++\sigma^-}{2}$ is
the averaged total scattering cross section~\cite{Siegmann:1994}).
Following this formula one can estimate the values for $\sigma^+$
and $\sigma^-$ in the MgO/Fe(110) bilayer using the experimentally
observed dependencies of $P$ on $J^+(d)$, $J^-(d)$ and $d$
yielding the values: $\sigma^+$=1.5\,nm$^{-1}$,
$\sigma^-$=5.3\,nm$^{-1}$ ($\sigma$=3.4\,nm$^{-1}$). The average
total scattering cross section value is clearly larger than the
value which can be expected for MgO. From the work of
Siegmann~\cite{Siegmann:1994} the value of about
$\sigma$=0.8\,nm$^{-1}$ for MgO (only $s$ and $p$ electrons in the
valence band) can be extracted. The scattering cross sections for
spin-up and spin-down electrons in MgO are different (different
$\vec{k}$ values for spin-up and spin-down electrons in the
valence band). On the basis of the observed significant difference
of the estimated value of $\sigma$=3.4\,nm$^{-1}$ and the expected
value of only $\sigma_{MgO}$=0.8\,nm$^{-1}$ one can suppose the
formation of, e.g., a thin FeO layer at the MgO/Fe interface. The
presence of such a FeO interfacial layer has been recently
experimentally identified by AES and STM in MgO/Fe(100) and
MgO/Fe(110) systems ~\cite{Vassent:1996,Fonin:2003}. Following
this assumption the spin polarization of emitted electrons can be
written as

$P=\frac{J^+_0\exp(-\sigma^+_{FeO}{\cdot}\,x)\exp(-\sigma^+_{MgO}{\cdot}\,d)-J^-_0\exp(-\sigma^-_{FeO}{\cdot}\,x)\exp(-\sigma^-_{MgO}{\cdot}\,d)}{J^+_0\exp(-\sigma^+_{FeO}{\cdot}\,x)\exp(-\sigma^+_{MgO}{\cdot}\,d)+J^-_0\exp(-\sigma^-_{FeO}{\cdot}\,x)\exp(-\sigma^-_{MgO}{\cdot}\,d)}$,

where $\sigma^+_{FeO}$ ($\sigma^-_{FeO}$) is the scattering cross
section for spin-up (down) electrons in the thin FeO interface
layer
($\sigma_{FeO}=\frac{\sigma^+_{FeO}+\sigma^-_{FeO}}{2}$=3.5\,nm$^{-1}$,
four holes in the valence band of
Fe$^{2+}$O$^{2-}$~\cite{Siegmann:1994}), $\sigma^+_{MgO}$
($\sigma^-_{MgO}$) is the scattering cross section for spin-up
(down) electrons in the MgO layer
($\sigma_{MgO}=\frac{\sigma^+_{MgO}+\sigma^-_{MgO}}{2}$=0.8\,nm$^{-1}$),
and $x$ is the thickness of the hypothetical FeO layer. From a
comparison of the two equations for $P$ the thickness of the FeO
interfacial layer ($x$) can be estimated by:

$x=d\cdot\frac{\sigma^++\sigma^--2\,\sigma_{MgO}}{2\,\sigma_{FeO}}$,

i.e. a linear dependence between the FeO interfacial layer
thickness ($x$) and the MgO overlayer thickness ($d$).

The calculated upper limit values of the FeO interfacial layer
thickness vs. MgO layer thickness are presented in the inset of
Fig.4. This model is suitable in case of abrupt interfaces in the
MgO/FeO/Fe(110) system and does not take into account scattering
at the interfaces, at possible Fe inclusions in the MgO layer, or
at defects etc. Thus the FeO layer thickness can actually be
smaller then estimated by our model. The formation and the
increase of the FeO layer thickness at the MgO/Fe interface with
increasing MgO layer thickness was also observed in recent AES
experiments performed at the MgO/Fe(110) system~\cite{Fonin:2003}.

In conclusion, the growth process of thin MgO films on Fe(110) and
the electronic structure of the MgO/Fe(110) interface have been
investigated. The SPARPES experiments show that the spin
polarization of the Fe(110) thin film strongly decreases with an
increasing MgO film thickness. This behavior can not be ascribed
to the scattering process of the spin-polarized photoelectrons in
only the MgO overlayer. In this case the formation of a thin iron
oxide layer at the MgO/Fe interface was supposed which describes
well the dependence of the spin polarization as function of the
MgO overlayer thickness. An upper limit of the iron oxide layer
thickness of two monolayers is deduced for an MgO layer of 7\,\AA.

This work was supported by BMBF through FKZ. 05KS1PAA/7.



\pagebreak

\begin{center}
\center {FIGURE AND TABLE CAPTIONS}
\end{center}

  {\bf Fig.1.} LEED images of (a) a 50\,{\AA} thick Fe(110) film on a W(110) substrate and
  (b) 30\,\AA\ of MgO on top of Fe(110) film. The energy of the primary electron beam
  is 123\,eV for (a) and 104\,eV for (b). (the sixth reflex in
  both pictures is not visible due to the sample holder).

  {\bf Fig.2.} PES spectra ($h\nu$=21.2\,eV) of the MgO/Fe(110) system as a function
  of the MgO layer thickness (shown on the right-hand side of
  the each spectra). The inset shows the change of the valence band edge (VBE) position
  of O (2\textit{p}) states with increasing MgO thickness
  for the MgO/Fe(110) system.

  {\bf Fig.3.} (a) The spin-resolved photoemission spectra (spin down: down triangle, spin up: up triangle)
  together with the total emission intensity (circles) for Fe(110),
  2\,\AA\,MgO/Fe(110), and 5\,\AA\,MgO/Fe(110) (from bottom to top). (b) The spin
  polarization as function of binding energy of a 50\,\AA\ thick Fe(110) film (solid square), 2\,\AA\,MgO/Fe(110)
  (open triangle up), and 5\,\AA\,MgO/Fe(110) (solid circle).

  {\bf Fig.4.} The change of the normalized spin polarization at
  E$_F$ of the MgO/Fe(110) system with an increasing MgO layer
  thickness. The spline fit to the experimental data is shown by a dot-dashed line.
  The inset presents the estimated thickness of the iron oxide layer at the MgO-Fe
  interface as a function of MgO layer thickness.

\end{document}